\documentclass[11pt]{article}
\title{An exact correspondence between Quantum Hall Skyrmions and
non linear $\sigma$-models }
\author{V. Pasquier\\
CEA/Saclay, Service de Physique Th\'eorique\\
F-91191 Gif-sur-Yvette Cedex, FRANCE
}

\newcommand{\zbar}{\overline z}
\newcommand{\del}{\partial}

\begin{document}
\maketitle
\abstract{ We show that the ground states of spin-fermions 
interacting
by a delta repulsive potential in a strong magnetic field
can be put in one to one correspondence with
the solitons of the $S_2$ non linear $\sigma$-model.   }

\section{Introduction}

The minimum of a classical action
can be related to the ground state of a quantum Hamiltonian.
In this letter we review the Skyrmion of the Hall effect \cite{SON}
which gives an example of such a correspondence 
and relates the topological charge of the classical soliton to the electric
charge of the quantum state.
In a given
topological sector, the solitons which minimize the action all have the same action
and are in one to one
correspondence with the degenerate eigenstates of a quantum Hamiltonian in the
same charge sector.
Moreover in both cases, the action and the energy are equal to the modulus of the charge. 

Belavin and Polyakov \cite{BEL} have considered the classical solutions of the two dimensional
nonlinear $\sigma$-model on the sphere $S_2$.
The field configurations $\vec n(x,y)$ can be characterized by their stereographic projection
on  the complex plane $w(x,y)$. One requires that the spin points in the $x$
direction at infinity, or equivalently $w(\infty)=1$.
The minima of the action
are rational fractions of $z=x+iy$: $w(z)=f(z)/g(z)$.  
The soliton 's winding number and its classical action are both given by the degree k
of the polynomials $f,g$. The soliton is thus determined by
the positions where the spin points to the south and the north pole
given by the zeros of $f$ and $g$. 

In quantum mechanics, the wave function for a single spin $1/2$ particle constrained
to the Lowest Landau Level (LLL) is fully determined by the positions
where the spin is up or down with probability one. 
Up to an exponential term, the two components of this wave function
are polynomials in $z$ vanishing at the positions where the spin is respectively up and down.
Such a wave function could for example describe a Bose condensate for two spices of particles
in a rotating frame.
This is not sufficient however for the expectation value of the spin 
operators to coincide with  $\vec n$
everywhere in space. To achieve this precisely one must consider manybody wave functions which
up to an exponential can be written as:
\begin{eqnarray}
\langle z_1,..,z_{N_e}|\Phi\rangle=\prod_{i=1}^{N_e} (f(z_i) \uparrow + g(z_i) \downarrow) \prod_{i<j} (z_i-z_j)^{m} 
\label {WAVE}
\end{eqnarray}
These wave functions have been considered (for $m=1$) by MacDonald, Fertig and Brey \cite{MAC}
in the Hall effect context. 
They represent the ground states of spin one half particles at a filling
fraction close to $1/m$.
Here we show 
that the correspondence with the nonlinear $\sigma$-model
can be made precise in the case $m=1$. 
One has the following correspondence table:
\smallskip

\centerline{
%\begin{table}[h]
\begin{tabular} {|c|c|}
\hline
$S=\int (\vec \nabla \vec n)^2 d^2x $&  $H=\sum_{i<j} \delta^{(2)} (\vec x_i-\vec x_j)$\\
\hline
$\vec n(\vec x)$ & $|\Phi\rangle {\rm \ a \ Slater\  determinant}$ \\
\hline
$S(\vec n)=k={\rm winding \ nb.}$ &$ \langle \Phi|H| \Phi \rangle=k={\rm electric\ charge}$\\
\hline
\end{tabular}}
%\end{table}

\section{Skyrmion and non linear $\sigma$-model}
\smallskip

Let us first review the classical solitons 
on the sphere $S_2$. A point on the sphere
is a unit vector $\vec n(\vec x)$ with which we construct the projector
$p(\vec x)=(1+\vec n\vec \sigma)/2$. $p$ is a two by two rank one projector $p^2=p$
and the action for the non linear $\sigma$-model takes the form:
\begin{eqnarray}
S={1\over \pi} \int {\rm tr}\ \partial_z p \partial_{\overline z} p \ d^2 x
\label {ACTION}
\end{eqnarray}
To obtain the solitons which minimize the action let us substitute $\partial_zp^2$ for $\partial_zp$ 
to rewrite the integrand as
${\rm tr}\ p  (\partial_z p \partial_{\overline z} p +\partial_{\zbar} p \partial_{z} p )$
and add to (\ref{ACTION}) the topological term:
\begin{eqnarray}
K={1\over \pi}\int {\rm tr}\ p  (\partial_z p \partial_{\overline z}p
- \partial_{\zbar} p \partial_{z} p)\  d^2 x
\label {ACTION2}
\end{eqnarray}
so that the sum takes the form:
\begin{eqnarray}
S'=S+K={2\over \pi}\int {\rm tr}\ (p\partial_{\overline z} p)^+(p\partial_{\overline z} p) \ d^2 x
\label {ACTION3}
\end{eqnarray}
(\ref{ACTION3}) is positive and the solutions with $S'=0$ must obey $p\partial_{\overline z} p=0$.
If we parameterize $p$ by a unitary vector $v$, $v^+v=1$, $p=vv^+$, it is solved for 
$v=N^{-1}(f(z),g(z))$ where $f,g$ are homomorphic functions and $N=\sqrt{|f|^2+|g|^2}$.
If one requires that $p(\infty)=(1+\sigma_x)/2$, $f$ and $g$ are polynomials with the same highest coefficient
$z^k$. The integrand of $K$ is the field strength of the gauge potential $\omega=-v^+dv/2i$
which goes to a pure gauge far from soliton. The topological term is therefore
given by the contour integral of $\omega$ at infinity equal to $-k$ 
and thus $S=k$.

The quantum analogous problem we consider here
consists in finding the degenerate ground states of 
electrons  interacting by 
a $\delta$ repulsive potential in the lowest Landau level (LLL).
The electrons are confined in a finite disc thread by $N_{\phi}$ magnetic fluxes.
When the number of electrons $N_e$ differs from  $N_{\phi}$
by an integer equal to the winding number $k$
the quantum eigenstates coincide with the classical 
solitonic field configurations
if the scale of variation of the soliton is large compared to the magnetic length.

In order to introduce the physical problem we need to define some notations \cite{PASLET}.
In the symmetric gauge the LLL degenerate wave functions can be taken as the eigenstates of the
angular momentum operator. They are given by:
\begin{eqnarray}
\langle z|l\rangle=(q^{1/2}z)^l/(2\pi l!)^{1/2}e^{-qz\zbar/2}
\label{BAS}
\end{eqnarray}
where the parameter $q$ has the dimension of a ${\rm length}^{-2}$.
Later we shall be interested in the limit $q\to \infty$ where we recover the classical limit.
We confine the electrons in a disc of area $\Omega=N_{\phi}\pi/q$ where $N_{\phi}$ is an
integer counting the number of magnetic fluxes through the disc. Thus we must
have $0\le l\le N_{\phi}-1$ and $N_{\phi}$ counts the degeneracy of the LLL.

It is useful to define a set of oscillators which act within the LLL:
\begin{equation}
b=\del_{z}/q+\zbar/2, \ b^+=-\del_{\zbar}/q +z/2 
\label{OSC}
\end{equation}
They obey the commutation relation: $[b,b^+]=q^{-1}$ and in the limit where $q\to \infty$ they
become the true coordinates of the electrons.
Things simplify if we take the size of the disc large
compared to the characteristic size of the solitons which amounts to take $N_{\phi}=\infty$.
In this limit
the coherent states $|z\rangle$ are defined as the eigenstates of $b$: $b|z\rangle=\zbar|z\rangle$
and their scalar product $\rho=\langle z|z\rangle=q/\pi$ does not depend on $|z\rangle$.
The Q-symbol \cite{KLAU} 
of a matrix $\hat A=\sum_{ll'}|l\rangle \hat A_{ll'}\langle l'|$
acting within the LLL consists in
bracketing it between coherent states and normalizing it by $\rho$:
\begin{eqnarray}
a(z,\zbar)&=&\langle z| \hat A|z\rangle/\rho
%e^{-p\bar p/2q}e^{i(\bar p z+p\zbar)}&=&\langle z| \hat e_{\vec p}|z\rangle \cr
\label{COMU}
\end{eqnarray}
In particular one has:
\begin{eqnarray}
e^{i(\bar p z+p\zbar)}
=\langle z|e^{i{\bar p b^+\over q}} e^{i {p b\over q}}|z\rangle / \rho
\label{COMU1}
\end{eqnarray}

The Q-symbol induces a non commutative product on functions which we denote by $*$:
$a*b=\langle z|\hat A \hat B|z\rangle/ \rho$ and 
in the limit $q\to \infty$, the *-product coincides with
the ordinary product. Using (\ref{COMU1}) one can evaluate the
first order correction to the ordinary product given by:
\begin{eqnarray}
a*b=ab+{1\over \pi \rho}\partial_{\bar z}a\partial_{ z}b +O({1\over \rho^{2}})
\label{COMU2}
\end{eqnarray}

The second quantized field that 
annihilates (creates) an electron with a spin $\sigma$
at position $\vec x$ in the LLL can be constructed
in terms of the fermionic operators $c_{l\sigma}\ (c^+_{l\sigma})$
which annihilate (create) an electron in the $l^{\rm th}$ orbital:
\begin{eqnarray}
\Psi_{\sigma}(\vec x)=\sum_l \langle z|l \rangle c_{l\sigma}
\label{FIELD}
\end{eqnarray}
In terms of this field, the total number of electrons in the LLL
is $N_e=\int\sum_{\sigma} \Psi^+_{\sigma}\Psi_{\sigma}(\vec x)\ d^2x$.
The charge of the Skyrmion is the difference between the number of magnetic fluxes $N_{\phi}$
and the number of electrons $N_e$: $Q_s=N_{\phi}-N_e$. In other words, it is the
number of electrons added or subtracted to the system starting from a situation
where the filling factor $\nu=N_e/N_{\phi}$ is exactly one.
In the following we consider the limit $N_{\phi},\ N_e=\infty$ keeping the charge $Q_s$ fixed.

The authors of \cite{MAC} observed that the zero energy
states of the hard-core model Hamiltonian could be completely determined.
In this letter, we consider
a closely related short range repulsive Hamiltonian 
invariant under a particle hole transformation $\Psi \to \Psi^+$
and such that the energy of its ground
state coincides with the charge.
It is given by:
\begin{eqnarray}
H={1\over \rho}\int(\Psi^+_{\uparrow}\Psi_{\uparrow}-\Psi_{\downarrow}\Psi^+_{\downarrow})^2(\vec x) \ d^2x
={2\over \rho}\int (\Psi_{\uparrow}\Psi_{\downarrow})^+(\Psi_{\uparrow}\Psi_{\downarrow})(\vec x) \ d^2x
+ Q_s
\label{HAMIL}
\end{eqnarray}
where we have used the fact that 
$\{\Psi_{\sigma}(\vec x),\Psi^+_{\sigma'}(\vec x)\}=\rho\delta_{\sigma\sigma'}$ 
to obtain the second equality.
Let us for simplicity consider the case where $Q_s>0$, the other case can be reached using
a particle hole transformation. 
This Hamiltonian is clearly bounded from bellow by $Q_s$ and the exact eigenstates with energy
$Q_s$ are obtained for states $|\Phi\rangle$ such that 
$\Psi_{\uparrow}(\vec x)\Psi_{\downarrow}(\vec x)|\Phi\rangle=0$.
In such a state two electrons never occupy the same position and the wave function is
blind to the short range potential. This property is precisely guaranteed by the factor
$\prod_{i<j}(z_i-z_j)$ in (\ref{WAVE}).

The states (\ref{WAVE}) carry a charge $Q_s=k$ where $k$ is the degree of the
polynomials $f$ and $g$. They are Slater determinants:
\begin{eqnarray}
\bigskip\langle z_1,..,z_{N_e}|
\Phi\rangle=\bigwedge_{1}^{N_e}(f(z_i)\uparrow+g(z_i)\downarrow)\langle z_i| \tilde l\rangle
\label{PHI}
\end{eqnarray}
In principle the states $|\tilde l\rangle$ can be the LLL basis (\ref{BAS})
but it is more convenient to redefine new states so that components of the antisymmetrized product
(\ref{PHI}) are
orthogonal to each other.
The basis $|\tilde l\rangle$ is an orthogonal basis for 
the scalar product: $\langle \tilde l |\bar f(b) f(b^+)+\bar g(b) g(b^+)|\tilde m\rangle$ and
$\langle z| \tilde l\rangle$ is a polynomial of degree $l$.
In other words, $\langle z| \tilde l\rangle$ is basis of orthogonal polynomials for the
scalar product:
\begin{eqnarray}
\langle \phi|\phi' \rangle=\int \bar \phi(\bar z)\phi'(z) e^{-q\bar z z}(|f|^2+|g|^2) d^2x
\end{eqnarray}

In general \cite{RIP}, a Slater determinant is fully determined by the matrix expectation value:
\begin{eqnarray}
\langle z|\hat P|z\rangle=
\rho p=
\langle \Phi|
\pmatrix{\Psi^+_{\downarrow}\Psi_{\downarrow}&\Psi^+_{\uparrow}\Psi_{\downarrow}\cr
\Psi^+_{\downarrow}\Psi_{\uparrow}&\Psi^+_{\uparrow}\Psi_{\uparrow}\cr}
|\Phi\rangle
%=\pmatrix{A&B\cr C&D}
\label{PROJ}
\end{eqnarray}
and its characteristic property is that $\hat P$ is a projector $\hat P^2=\hat P$.
In the case of (\ref{PHI}) we can obtain $\hat P$ explicitly as follows.
The states 
$|v_l\rangle=(f(b^+)\uparrow+g(b^+)\downarrow)|\tilde l\rangle$ can be organized
into a vector $V=\sum_l |v_l\rangle \langle l|$. 
By construction, $V^+V={\rm Id}$, so that, $VV^+=\hat P$, where $\hat P$
is the projector 
with Q-symbol $p$.

To relate the Skyrmion to the classical $\sigma$-model, let us evaluate the energy
of a slater determinant $|\Phi\rangle$ using the Wick theorem
\footnote{In \cite{PASLET} we had evaluated the energy keeping the Hartree contribution only.}:
\begin{eqnarray}
\langle \Phi|H|\Phi\rangle=\rho\int (2{\rm det}p-{\rm tr}p+1) \ d^2 x
\label{ENE}
\end{eqnarray}
In the above expression, the determinant is evaluated using the ordinary product.
Suppose we replace it with the $*$-product 
in (\ref{ENE}). Using the fact tat $p*p=p$ one verifies that the integrand rewrites
$({\rm tr}p-1)*({\rm tr}p-1).$  
Since ${\rm tr}(p-1)$ is ${\rm O}(\rho^{-1})$,
the *-square is
${\rm O}(\rho^{-2})$ and does not contribute to the
energy when $\rho \to \infty$. Therefore, the limiting value of (\ref{ENE})
is given by the modification induced by the ordinary product at first order in 
$\rho^{-1}$. One obtains from (\ref{COMU2}):
\begin{eqnarray}
\langle \Phi|H|\Phi\rangle=
{1\over \pi} \int {\rm tr}\ \partial_z p \partial_{\overline z} p \ d^2 x
+O(1/\rho)
\label{FIN}
\end{eqnarray}
which is the value of the action (\ref{ACTION}) and
establishes the correspondence between the classical and the quantum problems.

Although the classical action (\ref{ACTION}) 
can be obtained straightforwardly from the energy (\ref{ENE}) in the limit
$\rho \to \infty$,  
the topological term (\ref{ACTION2}) cannot so directly be
related to the charge of the Skyrmion.
It is nevertheless possible to define the topological term at the quantum level
\cite{CONNES}  and to verify it coincides with the charge in the present case.
For this we need to make the following substitutions in (\ref{ACTION2}):
\begin{eqnarray}
p \to \hat P,
\ \ \ {1 \over \pi} \int\ .\  d^2x \to {1 \over q}\rm{ Tr}\ .,
\ \ \ \partial_{z}. \to q[b,.],
\ \ \ \partial_{\zbar}. \to -q[b^+,.]
\end{eqnarray}
where $\rm{Tr}$ now stands for the trace of the matrix in the LLL Hilbert space.
The modified expression of $K$ (\ref{ACTION2}) still defines a topological invariant
which coincides with $K$ in the limit $\rho \to \infty$. One advantage is
that it can be defined for projectors $\hat P$ which do not have a classical limit $p$.
For example, the easiest way to realize a charge $-k$ configuration consists in expelling
$k$ electrons from the first $l<k$ angular momentum orbitals in the $\nu=1$ filled LLL
(Here the spin can be kept fixed and plays no essential role).
The projector which characterizes this configuration is 
$\hat P=\sum_{l\geq k} |l\rangle \langle l|$, equivalently 
$V=b^{+k} N^{-1}=\sum_{l}|l+k\rangle \langle l|$.
This projector has no classical correspondent because the size of the soliton
is of the order of the magnetic length. Nevertheless, using the quantum expression
one can easily verify that the topological invariant is equal to the charge $K=-k$.
One can also proceed as in the classical case and express K as the trace of a
commutator: $K=-q\rm{ Tr}([b,V^+[b^+,V]]-[b^+,V^+[b,V])$. Then use the fact that
the Hilbert space is finite dimensional 
(the angular momentum $l$ is bounded by $N_{\phi}$) to undo the commutator and
bring back the trace to a single  boundary ($l\sim N_{\phi}$) matrix element.

\section{Acknowledgments}
I thank D.Bernard, G.Misguich and D.Serban for discussions and 
G.Misguich for a careful reading of the manuscript.

\vfill\eject
\end{document}